\documentclass[aps,pra,twocolumn,superscriptaddress]{revtex4}
\usepackage{amsfonts}
\usepackage{amsmath}
\usepackage{amssymb}
\usepackage{graphicx}
\begin{document}
\title{\small Dynamics of kicked spin-orbit-coupled Bose-Einstein condensates}
\author{\footnotesize Qingbo Wang}
\affiliation{\footnotesize Key Laboratory for Microstructural Material Physics of Hebei Province, School of Science, Yanshan University, Qinhuangdao 066004, People's Republic of China}
\affiliation{\footnotesize School of Physics and Technology, Tangshan Normal University, Tangshan 063000, People's Republic of China}
\author{\footnotesize Wenjing Zhao}
\affiliation{\footnotesize Key Laboratory for Microstructural Material Physics of Hebei Province, School of Science, Yanshan University, Qinhuangdao 066004, People's Republic of China}
\author{\footnotesize Linghua Wen}
\email{linghuawen@ysu.edu.cn}
\affiliation{\footnotesize Key Laboratory for Microstructural Material Physics of Hebei Province, School of Science, Yanshan University, Qinhuangdao 066004, People's Republic of China}
\begin{abstract}
\footnotesize
We investigate the dynamics of kicked pseudo-spin-1/2 Bose-Einstein condensates (BECs) with spin-orbit coupling (SOC) in a tightly confined toroidal trap. The system exhibits different dynamical behaviors depending on the competition among SOC, kick strength, kick period and interatomic interaction. For weak kick strength, with the increase of SOC the density profiles of two components evolve from overlapped symmetric distributions into staggered antisymmetric distributions, and the evolution of energy experiences a transition from quasiperiodic motion to modulated quantum beating. For large kick strength, when the SOC strength increases, the overlapped symmetric density distributions become staggered irregular patterns, and the energy evolution undergoes a transition from quasiperiodic motion to dynamical localization. Furthermore, in the case of weak SOC, the increase of kick period leads to a transition of the system from quantum beating to Rabi oscillation, while for the case of strong SOC the system demonstrates complex quasiperiodic motion.

\end{abstract}
\maketitle
\footnotesize

The quantum $\delta$-kicked rotors (QKRs) in Bose-Einstein condensates (BECs) have received much attention in recent decades. Since the first experiment of the QKR on ultracold atoms in 1995 \cite{Moore1995,Sadgrove2011}, succeeding QKR studies have mostly employed kicked BECs due to the need for well-defined initial momenta, and revealed a rich variety of effects including quantum resonances \cite{Moore1994,Fishman2002,Ryu2006,Talukdar2010}, quantum accelerator modes \cite{Iomin2002,Behinaein2006}, quantum ratchets \cite{Sadgrove2007,White2013,Ni2016,Hainaut2018}, quantum walks \cite{Wei2015,Summy2016,Dadras2018}, and dynamical localization \cite{Marino2019}. For a kicked BEC, strong interatomic interaction will result in an instability of the condensate, where the transition to instability might be associated with a transition to quantum chaos or with resonant driving of Bogoliubov modes \cite{Zhang2004,Liu2006,Reslen}. Recently, subdiffusive behavior has been predicted in the long-time dynamics of a kicked scalar BEC with weak enough interactions \cite{Lellouch2020}. For a quasiperiodic kicked rotor, Bogoliubov excitations can give rise to a quasi--insulator-to-metal transition \cite{Vermersch}. In addition, Zhou \emph{et al} explored the Floquet topological phases in a double kicked rotor \cite{Zhou}. To the best of our knowledge, most of the existing investigations concerning on the kicked BECs as mentioned above focus on various dynamics regimes of a single-component BEC or different types of kicked rotors. However, there are few studies so far on the dynamical behaviors of kicked two-component BECs with spin-orbit coupling (SOC).

In fact, SOC describes the interaction between the spin and the momentum of a quantum particle, which is one of the key factors in determining the physics of the BEC system. SOC sustains a unique dispersion relationship, and the competition between dispersion relationship and contact interaction can lead to many novel quantum phases and intriguing physical features \cite{Zhai2015,Wang2020,Wang2021}. Recently, the synthetic one-dimensional (1D) SOC, 2D SOC and 3D SOC in BECs have been realized experimentally \cite{Lin2011,Wu2016,Qu2013,WangZY}, which provides a brand-new platform for the quantum simulation of condensed matter physics and also for the exploration of exotic properties of multi-component BECs usually unaccessible in scalar BECs and electronic materials. In this context, it is of particular interest to investigate the dynamical properties of kicked interacting BECs with SOC.

We consider a 1D system of kicked pseudo-spin-1/2 BECs with SOC in a tightly confined toroidal trap (the radius $R$ and thickness $r$ of the trap satisfy $R\gg r$). Under the periodic boundary condition, the system can be equivalent to a 1D system along the $x$-axis. The model Hamiltonian is given by
\begin{equation}
H=K\cos xf(t)+H_0+H_I,
\end{equation}
where $K$ is the kick strength. The periodic QKR is described by
\begin{equation}
f(t)=\sum^{N_k}_{n=1} \delta(t-nT),
\end{equation}
where $T$ and $N_k$ are the kick period and the total number of kicks. $T$ is set to a rational multiple of $\pi$ \cite{Fishman2002}. The single particle Hamiltonian \cite{Lin2011,Qu2013} reads
\begin{eqnarray}
H_0=\frac{\hbar^2{k}^2_x}{2m}I+\frac{\Omega}{2}\sigma_x+\frac{\delta_L}{2}\sigma_z+2\alpha k_x\sigma_z+E_LI,
\end{eqnarray}
where $k_x$ is the quasimomentum, and $m$ is the atomic mass. $\sigma_{x}$ and $\sigma_{z}$ are Pauli operators. $\Omega$ and $\delta_L$ are the Raman coupling strength and detuning. In the present work, we assume $\Omega=0$ and $\delta_L=0$. $\alpha=E_L/k_L=\hbar^2k_L/2m \propto k_L $ is the strength of SOC. $E_L=\hbar^2k_L^2/2m$ and $\hbar k_L=\sqrt{2}\pi\hbar/\lambda$ are the recoil energy and momentum, which depend on the Raman laser with wavelength $\lambda$. Then, $H_0$ can be written as
\begin{eqnarray}
H_0=\frac{\hbar^2}{2m}\begin{pmatrix}(k_x+k_L)^2 & 0 \\ 0 &(k_x-k_L)^2\end{pmatrix},
\end{eqnarray}
where we have omitted the irrelevant constant terms. The contact interaction is described by
\begin{eqnarray}
H_I=\begin{pmatrix}g_{1}\left\vert\psi_1\right\vert ^{2}+g_{12}\left\vert\psi_2\right\vert ^{2} & 0 \\ 0 &g_{21}\left\vert\psi_1\right\vert ^{2}+g_{2}\left\vert\psi_2\right\vert ^{2}\end{pmatrix}.
\end{eqnarray}
The 1D coefficients are given by $g_{j}=2 a_{j}N/a_{\perp}^2$($j$=1,2) and $g_{12}=g_{21}=2 a_{12}N/a_{\perp}^2$, where $a_{j}$ and $a_{12}$ are the s-wave scattering lengths between intra- and inter-component atoms, $N$ is the number of atoms, and $a_{\perp}=\sqrt{\hbar/m\omega_{\perp}}$ with $\omega_{\perp}$ the radial trap frequency. For simplicity, we assume $g_{j}=g_{12}=g$ throughout this work. By introducing the dimensionless parameters via notations $E_0=\hbar^2k_0^2/m$, $\widetilde{H}_0=H_0/E_0$, $k_0=\sqrt{2}\pi/\lambda$, $\widetilde{k}_x=k_x/k_0$, $\widetilde{K}=K/E_0$, $\widetilde{x}=x/{k_0}$, $\widetilde{t}=2E_0t/\hbar$, $\widetilde{g}=g/E_0$, and $\gamma=k_L/k_0$ (the dimensionless strength of SOC), Eq. (1) can be expressed as
\begin{eqnarray}
H&=&K\cos xf(t)+\mathrm{diag}\big(\frac{1}{2}(k_x+\gamma)^2,\frac{1}{2}(k_x-\gamma)^2\big)\notag \\&&+g\;\mathrm{diag}\big(\left\vert\psi_1\right\vert ^{2}+\left\vert\psi_2\right\vert ^{2},\left\vert\psi_1\right\vert ^{2}+\left\vert\psi_2\right\vert ^{2}\big),
\end{eqnarray}
where the tildes are omitted for simplicity. The average energy per atom $\langle E\rangle$ is given by
\begin{eqnarray}
\langle E\rangle&=&\int_{-\pi}^{\pi}dx\biggl(\frac{1}{2}(k_x+\gamma)^2\left\vert\psi_1\right\vert ^{2}+\frac{1}{2}(k_x-\gamma)^2\left\vert\psi_2\right\vert ^{2}\notag \\&&
+\frac{1}{2}g(\left\vert\psi_1\right\vert ^{4}+\left\vert\psi_2\right\vert ^{4})+g\left\vert\psi_1\right\vert ^{2}\left\vert\psi_2\right\vert ^{2}\biggr).
\end{eqnarray}
The dynamic equation of the system can be expressed as
\begin{eqnarray}
i\frac{\partial}{\partial t}\begin{pmatrix}\psi_1 \\\psi_2 \end{pmatrix}=H\begin{pmatrix}\psi_1 \\\psi_2 \end{pmatrix},
\end{eqnarray}
with the initial wave function $\psi_j(x,0)=1/{\sqrt{4\pi}}$ ($j$=1,2). The system satisfies the normalization condition $\int_{-\pi}^{\pi}\bigl(\left\vert\psi_1\right\vert ^{2}+\left\vert\psi_2\right\vert ^{2}\bigr)dx=1$ and the periodic boundary condition $\psi_j(x,t)=\psi_j(x+2\pi,t)$.
\begin{figure}[htpb]
\centerline{\includegraphics*[width=8.5 cm]{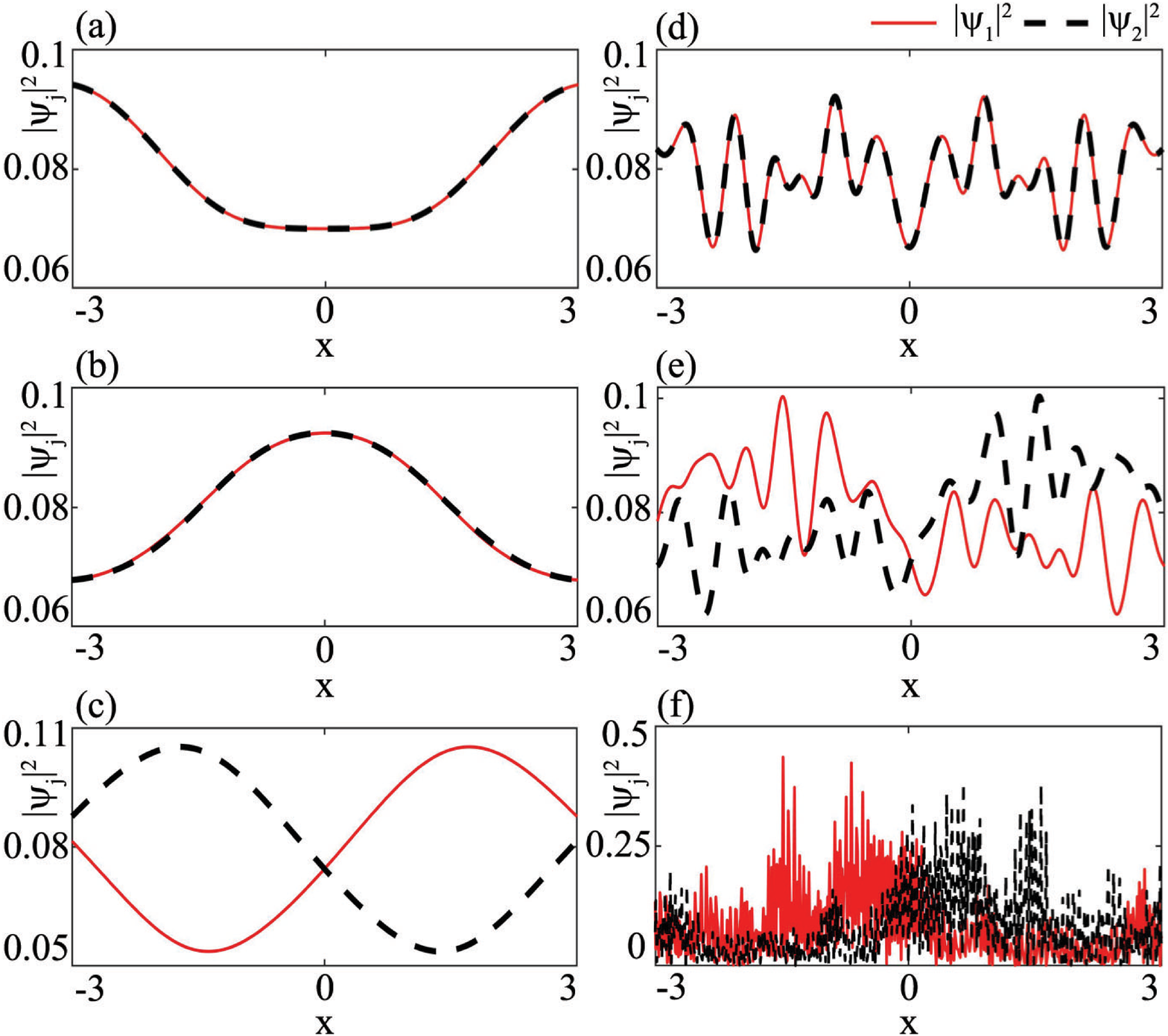}}
\caption{\scriptsize The density distributions for $N_k$=800, $T$=2$\pi$, and $g$=1. (a)-(c): $\gamma$=0, 10, 40 and $K$=0.8; (d)-(f): $\gamma$=0, 10, 40 and $K$=8, respectively.}
\end{figure}

Figure 1 shows the effects of SOC and kick strength on the density distributions of the system for $N_k$=800, $T$=2$\pi$, and $g$=1. One can see the remarkable differences among the cases of zero (weak) SOC (the first row), moderate SOC (the second row), strong SOC (the third row), small kick strength (the left column), and large kick strength (the right column). For the case of small kick strength, with the increase of SOC, the density distributions of two components undergo a sequence of transitions: from overlapped symmetric distributions to basically overlapped symmetric distributions and then to staggered antisymmetric distributions. This is mainly due to the unique dispersion relationship of SOC \cite{Zhai2015,Lin2011}, which effectively affects the symmetry of the density distribution of the system, and SOC gradually occupies a dominant position in the competition between contact interaction and SOC. However, for the case of large kick strength, the two components exhibit overlapped symmetric distributions, staggered asymmetric distributions, and complex irregular distributions, respectively. The reason can be simply attributed to the large kick strength and the increasing SOC strength. Physically, SOC means the coupling between the internal states and the orbit motion of the atoms, and it breaks the parity symmetry and SU(2) symmetry of the system. The larger the kick strength and the SOC strength are, the more significant the SOC effect becomes, which implies that the quantum interference of the system can be enhanced or weakened depending on the initial conditions and the interplay of multiple parameters.

\begin{figure}[htpb]
\centerline{\includegraphics*[width=8.4 cm]{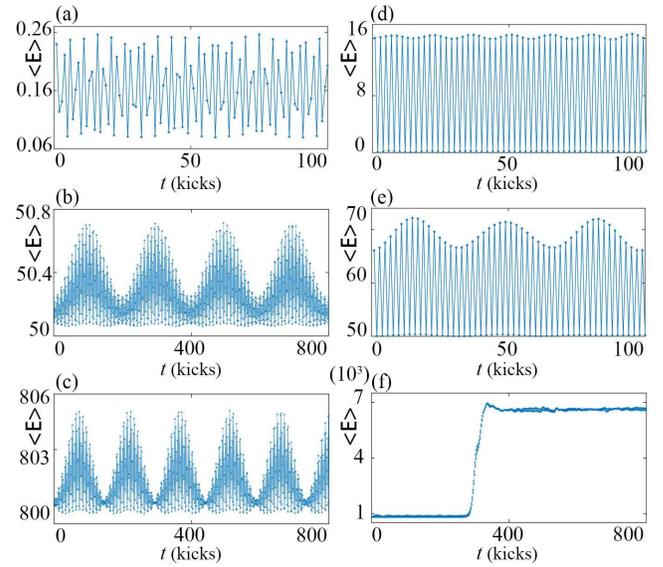}}
\caption{\scriptsize The evolution dependence on the average energy per atom $\langle E\rangle$ and the number of kicks $t$ for $N_k$=800, $T$=2$\pi$, and $g$=1. (a)-(c): $\gamma$=0, 10, 40 and $K$=0.8; (d)-(f): $\gamma$=0, 10, 40 and $K$=8, respectively.}
\end{figure}
\begin{figure}[htpb]
\centerline{\includegraphics*[width=8.4 cm]{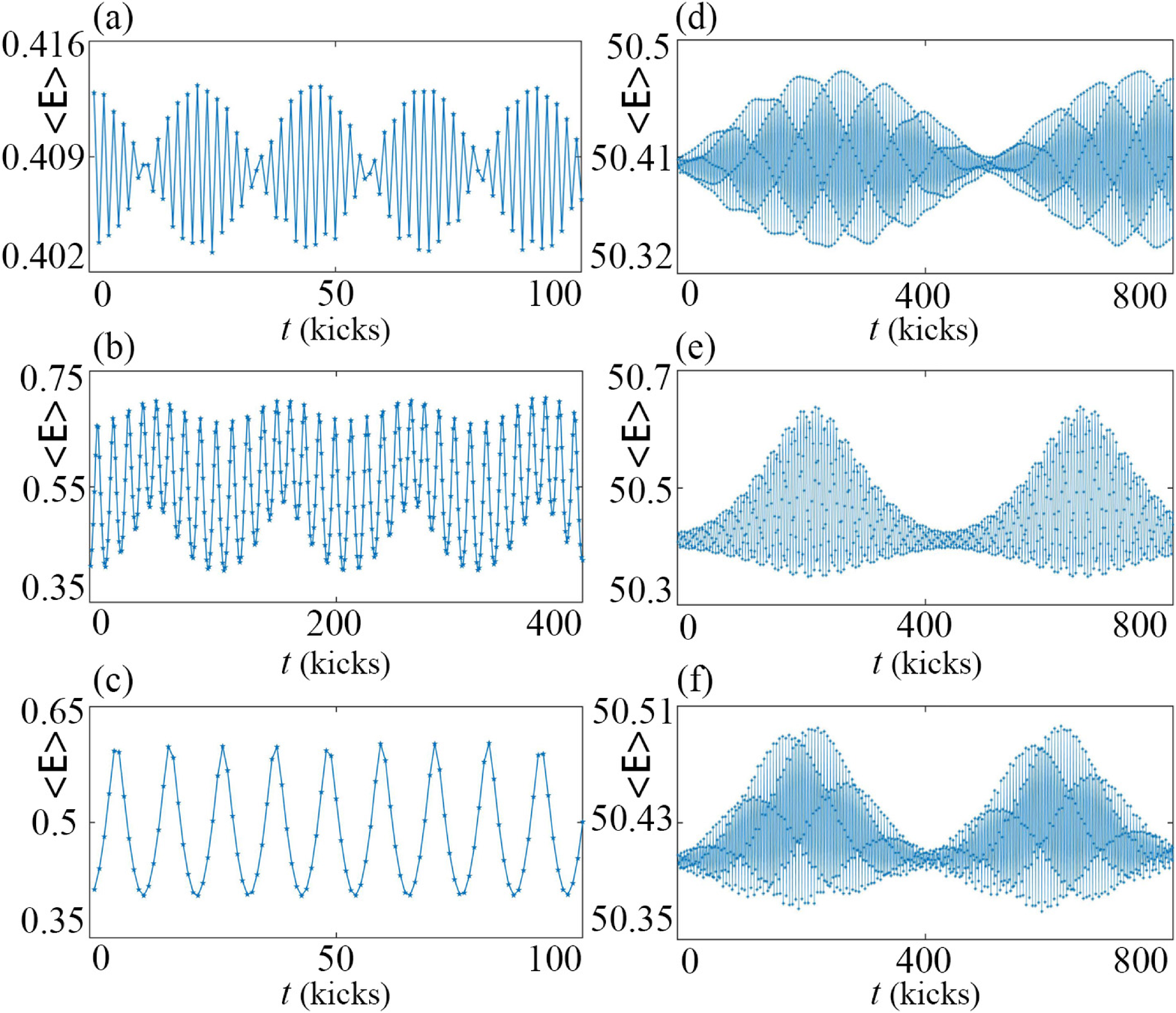}}
\caption{\scriptsize The evolution dependence on the average energy per atom $\langle E\rangle$ and the number of kicks $t$ for $N_k$=800, $K$=0.2, and $g$=5. (a)-(c): $\gamma$=0.1 and $T$=$\pi$, 2$\pi$, 4$\pi$; (d)-(f): $\gamma$=10 and $T$=$\pi$, 2$\pi$, 4$\pi$, respectively.}
\end{figure}

The dynamic process can also be characterized by the evolution of the average energy per atom $\langle E\rangle$ in Fig. 2, where the parameter values in Figs. 2(a)-(f) correspond to those in Figs. 1(a)-(f), respectively. For a single-component BEC in the absence of interatomic interactions, the typical dynamics of the system is periodic motion (quantum antiresonance) as pointed in \cite{Sadgrove2011,Zhang2004,Liu2006}, where the energy oscillates between two values. For the present system, once the contact interaction is included, the system exhibits typical characteristics of quasiperiodic motion as a result of the energy oscillation being modulated by contact interaction [Fig. 2(a)]. For the case of small kick strength, when the SOC strength increases from 0 to 10 and then to 40, the system exhibits novel variants of quantum beating, where the fluctuation at the bottom of the energy evolution curve is very small [Figs. 2(b)-(c)]. The modulated quantum beatings are remarkably different from the standard ones in a kicked BEC without SOC \cite{Zhang2004,Liu2006}, which indicates that SOC plays a key role in determining the quantum dynamics of kicked BECs. In addition, the maximum energy amplitude is proportional to the SOC strength, while the revival period is inversely proportional to the SOC strength. For the case of large kick strength and without SOC, the system shows an approximate quantum antiresonance behavior [Figs. 2(d)] in which the fluctuation at the top of the energy evolution curve is almost negligible (the oscillation amplitude approaches a constant value). This feature is owing to the quantum nature of the kicked BECs and the large kick strength. As is well known, for a classical kicked rotor, increasing kick strength will destroy the regular periodic or quasiperiodic motions of the rotor and will result in the transition to chaotic motions, featured by the diffusive growth of the energy \cite{Sadgrove2011,Liu2006}. In the absence of SOC, when the kick strength is much greater than the interaction strength, the system is approximately equivalent to a noninteracting quantum kicked rotor, which leads to such a similar antiresonance phenomenon. When the SOC strength increases to 10, the quasiperiodic motion becomes distinct as demonstrated in Fig. 2(c), where the upper value of the oscillation amplitude shows periodic oscillation. In particular, when $\gamma$=40 the energy has a sudden jump at $t=280$ and then exhibits a transition to dynamical localization, characterized by the quantum suppression of diffusive growth in energy \cite{Ringot2000}. Essentially, the dynamical localization is originated from the subtle quantum interference, featured by the fact that the wave function acquires the same phase for each kick, i.e, the effect of kicks adds coherently. Different from previous studies \cite{Hainaut2018,Marino2019}, here the dynamical localization is primarily generated by the interplay of strong SOC and large kick strength.

In Fig. 3, we show the effects of SOC and kick period on the dynamics of the system for $N_k$=800, $K$=0.2, and $g$=5. Here we take a relatively strong contact interaction, which makes the system more sensitive to the change of the kick period \cite{Marino2019}. For very weak SOC strength, when $T=\pi$, the amplitude of the oscillation decreases gradually to an approximate constant value and then revives [Fig. 3(a)]. This dynamic behavior is a typical quantum beating. Similar quantum beating phenomenon has also been observed in a kicked single-component BEC \cite{Zhang2004,Liu2006}, but here the parameter conditions (e.g., $T=\pi$ and relatively strong nonlinear interaction), physical system (spin-1/2 BECs with SOC) and formation mechanism are obviously different from those in \cite{Zhang2004,Liu2006}. With the increase of kick period (integral multiple of $\pi$), the system gradually evolves from standard quantum beating to quasiperiodic motion and then to Rabi oscillation [Figs. 3(a)-(c)]. During the transition, the system still keeps quasiperiodic or periodic motion despite the different patterns. For large SOC strength and various kick periods, the system displays different quasiperiodic motions, and the energy evolution patterns become rather complicated [Figs. 3(d)-(f)], where the revival periods are much larger than those in the weak SOC case. Therefore, from the perspective of quantitative analysis in the future experiments, we suggest that the SOC strength and the interaction strength should not be adjusted too much at the same time, so that the relatively simple dynamics in the system can be well investigated and tested.

In summary, we have studied the dynamics of kicked two-component BECs with SOC in a tightly confined toroidal trap. The combined effects of SOC, kick strength, kick period and contact interaction on the component density distributions and the energy evolution of the system are discussed systematically. The present system sustains rich dynamical behaviors, such as quantum antiresonance, standard quantum beating and modulated quantum beating, dynamical localization, Rabi oscillation, conventional quasiperiodic motion, and unusual complex quasiperiodic motion. In the case of small kick strength, when the SOC strength increases, the component density distributions change from overlapped symmetric profiles into staggered antisymmetric profiles. At the same time, the system undergoes a transition from quasiperiodic motion to modulated quantum beating. In the case of large kick strength, with the increase of SOC strength, the component density profiles evolve from overlapped symmetric distributions into irregular complex distributions. Correspondingly, the system experiences a transition from quantum antiresonance to dynamical localization. In addition, the effects of SOC strength and kick period on the energy evolution are analyzed. For weak SOC, the increase of kick period leads to a transition of the system from quantum beating to Rabi oscillation, while for the case of strong SOC the system demonstrates complex quasiperiodic motion. Our findings provide new understandings for the physical properties of kicked BECs.

\noindent{\textbf{\scriptsize CRediT authorship contribution statement}}

\scriptsize
\textbf{Qingbo Wang:} Formal analysis, Data curation, Writing-original draft. \textbf{Wenjing Zhao:} Formal analysis, Data curation. \textbf{Linghua Wen:} Conceptualization, Supervision, Formal analysis, Writing-review \& editing.

\noindent{\textbf{\scriptsize Acknowledgments}}

\scriptsize
We thanks Prof. Chuanwei Zhang and Prof. Chunlei Qu for helpful discussions. This work was supported by the National Natural Science Foundation of China (Grant Nos. 11475144 and 11047033), the Natural Science Foundation of Hebei Province, China (Grant Nos. A2019203049 and A2015203037), Science and Technology Innovation Ability Cultivation Foundation of Hebei Province, China (For college students), and Research Foundation of Yanshan University, China (Grant No. B846).

\end{document}